\pdfoutput=1
\documentclass[onecolumn,a4paper,fleqn,usenatbib]{mnras}

\usepackage[T1]{fontenc}
\usepackage{ae,aecompl}

\usepackage{graphicx}	
\usepackage{amsmath}	
\usepackage{amssymb}	

\usepackage{newtxtext,newtxmath}
\usepackage[T1]{fontenc}
\usepackage{ae,aecompl}

\usepackage{graphicx}	
\usepackage{amsmath}	

\usepackage{deluxetable}

\renewcommand {\deg}   {\mbox{$^\circ$}}

\newcommand   {\kms}   {\mbox{km\,s$^{-1}$}}
\renewcommand {\ga}    {\mbox{\rlap{\hbox{\lower5pt\hbox{$\sim$}}}\hbox{$>$}}}
\renewcommand {\la}    {\mbox{\rlap{\hbox{\lower5pt\hbox{$\sim$}}}\hbox{$<$}}}

\title[Kinks]{G359.13142-0.20005: A steep spectrum radio pulsar candidate with an X-ray counterpart running into the Galactic Center Snake 
(G359.1-0.2)}

\author[F. Yusef-Zadeh, Jun-Hui Zhao, R. Arendt, et al.]
{F. Yusef-Zadeh$^1$\thanks{E-mail: zadeh@northwestern.edu}, Jun-Hui Zhao$^2$, 
R. Arendt$^3$, M. Wardle$^{4}$, C. O. Heinke$^5$, M. Royster$^6$, 
C. Lang$^7$, 
\& J. Michail$^1$\\
$^{1}$Department of Physics and Astronomy Northwestern University, Evanston, IL 60208\\
$^{2}$Center for Astrophysics | Harvard-Smithsonian, 60 Garden Street, Cambridge, MA 02138, USA\\
$^{3}$UMBC/GSFC/CRESST 2, Code 665, NASA/GSFC, 8800 Greenbelt Rd, Greenbelt MD 20771\\
$^{4}$School of Mathematical and Physical Sciences,  Centre for Astronomy and \\
Space Technology, Macquarie University, Sydney NSW 2109, Australia\\
$^{5}$Department of Physics, CCIS 4-183, University of Alberta, Edmonton, AB T6G 2E1, Canada\\
$^{6}$Department of Physics, College of the Sequoias,  CA, 93277, USA\\
$^{7}$Department of Physics and Astronomy, University of Iowa, Iowa City, IA, 52242, USA\\
}
\date{Accepted XXX. Received YYY; in original form ZZZ}

\pubyear{2019}
\begin{document}
\label{firstpage}
\pagerange{\pageref{firstpage}--\pageref{lastpage}}
\maketitle

\def\msol{\hbox{$\hbox{M}_\odot$}}
\def\lsol{\hbox{$\hbox{L}_\odot$}}
\def\kms{km s$^{-1}$}
\def\Blos{B$_{\rm los}$}
\def\etal   {{\it et al.}}                     
\def\psec           {$.\negthinspace^{s}$}
\def\pasec          {$.\negthinspace^{\prime\prime}$}
\def\pdeg           {$.\kern-.25em ^{^\circ}$}
\def\degree{\ifmmode{^\circ} \else{$^\circ$}\fi}
\def\ut #1 #2 { \, \textrm{#1}^{#2}} 
\def\u #1 { \, \textrm{#1}}          
\def\nH {n_\mathrm{H}}
\def\ddeg   {\hbox{$.\!\!^\circ$}}              
\def\deg    {$^{\circ}$}                        
\def\le     {$\leq$}                            
\def\sec    {$^{\rm s}$}                        
\def\msol   {\hbox{$M_\odot$}}                  
\def\i      {\hbox{\it I}}                      
\def\v      {\hbox{\it V}}                      
\def\dasec  {\hbox{$.\!\!^{\prime\prime}$}}     
\def\asec   {$^{\prime\prime}$}                 
\def\dasec  {\hbox{$.\!\!^{\prime\prime}$}}     
\def\dsec   {\hbox{$.\!\!^{\rm s}$}}            
\def\min    {$^{\rm m}$}                        
\def\hour   {$^{\rm h}$}                        
\def\amin   {$^{\prime}$}                       
\def\lsol{\, \hbox{$\hbox{L}_\odot$}}
\def\sec    {$^{\rm s}$}                        
\def\etal   {{\it et al.}}                     
\def\la{\lower.4ex\hbox{$\;\buildrel <\over{\scriptstyle\sim}\;$}}
\def\ga{\lower.4ex\hbox{$\;\buildrel >\over{\scriptstyle\sim}\;$}}

\begin{abstract} 
The Snake is a remarkable Galactic center radio filament with a morphology characterized by two kinks along its $\sim 20'$ extent.  The 
major and minor kinks are located where the filament is  most distorted from a linear magnetized structure running perpendicular to the 
Galactic plane. We present {\em Chandra}, VLA, and MeerKAT data and report the detection of an  X-ray and radio source at the 
location of the major  kink. 
High-resolution radio images of the major kink reveal 
a  compact source with a steep spectrum with spectral index 
$\alpha\sim-2.7$ surrounded by extended emission.  
The radio luminosity and steep spectrum  of the compact source   are  consistent with 
a pulsar. 
We also show flattening of the spectrum and enhanced synchrotron emissivity away from the position of the 
major  kink along the Snake, which suggests injection of relativistic particles along the 
Snake. We argue that the major kink 
is created by 
a fast-moving ($\sim500-1000$ km s$^{-1}$), object
punching into the Snake, distorting its  magnetic structure, and producing  X-ray emission.  
X-ray emission pinpoints  an active acceleration site  where the interaction is taking place. 
A secondary   kink is argued to be induced by the impact of the high-velocity object producing the major kink. 
\end{abstract}


\begin{keywords}
cosmic rays: Interstellar Medium - 
ISM: magnetic fields - 
stars: pulsars - 
Galaxy: centre
\end{keywords}


\section{Introduction} 

It has been close to 40 years since the discovery of the nonthermal radio filaments (NRFs) associated with the Radio Arc in the Galactic 
Center \citep{zadeh84}.  NRFs with similar characteristics  have been discovered in the intervening years with a 
range of  morphology. 
 Recently, radio continuum MeerKAT images of 
the inner $3.5^\circ\times2.5^\circ$ of the Galactic 
center at 20 cm revealed  an order of magnitude increase in their number \citep{heywood19}. NRFs appear  as single filaments 
or groups of parallel filaments  with mean angular spacing  $\sim16''$ \citep{zadeh22a}. The mean spectral index of all 
the filaments ($\alpha\sim-0.83$, where 
$\alpha = d(\log F_\nu) / d(\log\nu)$,  
is steeper than that of supernova remnants 
(SNRs; $\alpha = -0.62$). The equipartition magnetic field strengths along the filaments range from $\sim100$ to 0.4~$m$G depending on the 
assumed ratio of cosmic-ray protons to electrons. NRFs run mainly perpendicular to the Galactic plane. 
The intrinsic polarization observed from these filaments shows that their internal magnetic fields are directed along the filaments 
\citep{zadeh97,pare19}. The mechanisms responsible for accelerating particles to relativistic energies and for creating the elongated 
filamentary geometry are still mysterious. Some authors have examined how the filaments might arise from their interaction with molecular 
and ionized clouds or with mass-losing stars \citep{rosner96,shore99,bicknell01,zadeh19}. Although numerous other models have been 
proposed to explain the origin of the filaments, there is no consensus how these filaments are produced 
\citep{nicholls95,rosner96,shore99,bicknell01,dahlburg02,zadeh03,boldyrev06,ferriere09,zadeh19,thomas20,sofue20,coughlin21,gopal24}. Broadly 
speaking nonthermal filament (NTF) models invoke either a single  energetic event at the Galactic center,  
a compact source accelerating cosmic 
rays to high energies, or a process that involves the turbulent environment of the Galactic center hosting a high cosmic-ray flux.
A similar mystery to the origin of the filaments  is the recent discovery of long structures in radio galaxies \citep{rudnick22}. 
Recent  comparison of the  populations of magnetized filaments in the Galactic center  and radio galaxies in the intercluster medium (ICM) suggests 
that that they are analogous to each other as might be anticipated based on their similar morphologies, 
in spite of vastly different physical properties, such as lengths, widths and magnetic field strengths \citep{zadeh22b}.

The Snake is one of the longest,  $\sim30'$ ($\sim70$ pc), 
brightest,  and most spectacular filaments in the Galactic 
center  distinguishing  itself by the presence of two  kinks along its length
 \citep{gray91,gray95}. \ion{H}{1} 
absorption line studies imply that the Snake is located at the 8 kpc distance of the Galactic center 
\citep{uchida92}. 
This filament is narrow ($<10''$ or $<0.4$ pc in width) and runs perpendicular to the Galactic plane \citep{heywood19}. Like numerous 
other filaments in the Galactic center, the Snake shows no obvious external powering source. Most Galactic center filaments exhibit 
smooth curvature, but  the Snake shows two distorted kinks, north (major kink)  and south (minor kink), 
and three different curvatures along its length 
\citep{gray95,zadeh22a}. 
It was also reported that two compact radio sources,  G359.132-0.200 and G359.120-0.265, are  located to the west of the major and minor  
kinks, respectively. The equipartition magnetic field of the Snake is strongest,  $\sim0.15$ mG,  to its north as it decreases to its south 
\citep{zadeh22a}. Faraday rotation measures (RMs) of 5500 and 1400 rad m$^{-2}$ have been reported to be due to an external and internal medium, 
respectively \citep{gray95}.

Chandra, XMM and NuStar have detected X-ray emission from a handful of radio filaments with X-ray spectral indices consistent with 
synchrotron emission. 
The X-ray spectral index and normalization is consistent with a broken power-law 
spectrum \citep{sakano03,lu03,zadeh05,zhang14,ponti15,zadeh21}. Here we report 
the  detection of X-ray emission 
from a source adjacent to the Snake, near the location of the 
major  kink. There is a compact radio  counterpart to the X-ray source at 6 and 1.2 GHz with a steep spectrum. 
Both radio and  X-ray emission appear to show extended structure pointing in the direction of the Snake. 
We suggest  a picture in which the major  and minor  kinks are  created, likely  by a fast-moving pulsar, 
 running into the Snake, and distorting 
the magnetic structure of the Snake. 
This  interaction picture 
is supported by the following  measurements: a spectral change in the radio along the 
filament,  a change in  the morphology, and  X-ray emission.

\begin{figure}
\centering
\includegraphics[scale=0.67,angle=0]{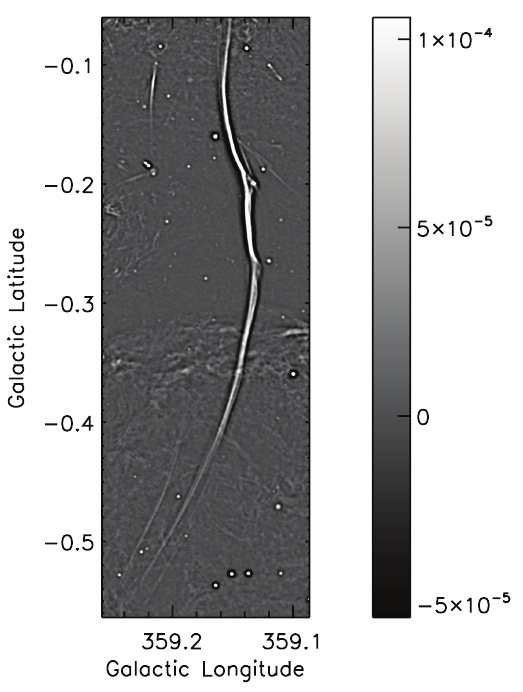}
\includegraphics[scale=0.67,angle=0]{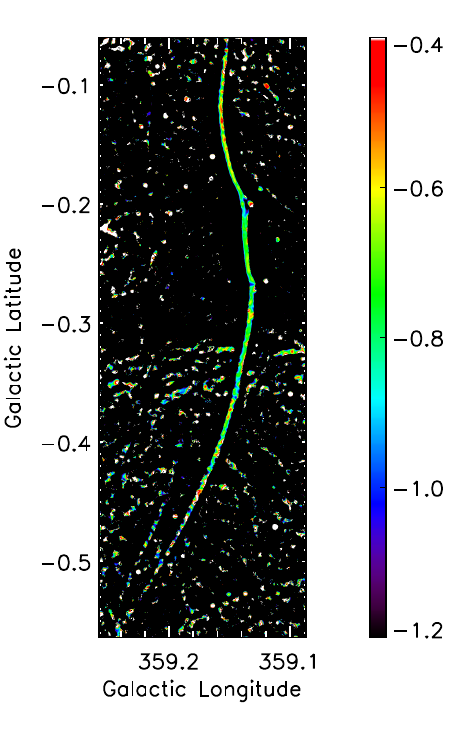}
\includegraphics[scale=0.65,angle=0]{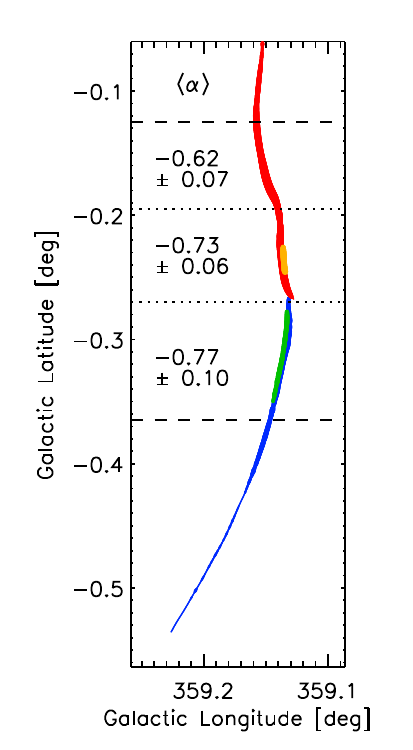}
\caption{
{\it (Left, a)}
A filtered image of the  the total intensity of the Snake at 1.28 GHz with $6.4''$ resolution \citep{zadeh22a} based on MeerKAT observations 
\citep{heywood22}. 
The unit in the grey scale bar is Jansky beam$^{-1}$. 
{\it (Middle, b)} 
The in-band  spectral index  at 1.28 GHz (20cm or L band) values along the Snake stretched to values between -0.83 and -0.31.  
The colors are used as key to identify the 4 sub-filaments (see  Fig. 2). 
The spectral index  in the color scale bar ranges between -1.2 and -0.4. 
{\it (Right, c)} 
Color traces  of different sub-filaments in the Snake. 
The colors are arbitrarily chosen to distinguish the four sub-filaments. The mean spectral indices within three bright intervals of the Snake are 
listed (and shown in more detail in Figure 2). The uncertainties on $<\alpha>$  are the standard deviation of $\alpha$ for each segment.  
}
\end{figure}


\section{Observations}



\subsection{Radio VLA data}
\pdfoutput=1

\subsubsection{X-band (wide) data} 

As part of a  targeted  survey of the central few   degrees of the Galactic center, 
six overlapping  fields were observed  toward the Snake using  the VLA X-band (wide) receiver at 10 GHz  
in the C-array configuration  on July 31 and August 16, 2021. 
The correlator was configured for 64 channels in each of  32 sub-bands
with a channel width of 2 MHz, producing 
producing all four Stokes parameters (RR, RL, LR and LL) with a total bandwidth of 4 GHz covering a frequency
range from 8 to 12 GHz. 3C 286, NRAO 530 and J1744-3116 were interleaved in the the survey observing runs for calibrations
of complex gains, bandpass, and flux density scales. 
The initial calibrations were carried out using a VLA pipeline program. 
The residual corrections were further  conducted using a model of J1744-3116 including both the core and extended structure.  
Table 1 summarizes the observations of the six mosaic fields and lists  the images produced from the observations of the  Snake.
Polarization analysis of the Snake, providing the intrinsic direction of the magnetic field and the rotation measure distribution,  will be given elsewhere.



\begin{figure}
\centering
\includegraphics[scale=0.85,angle=0]{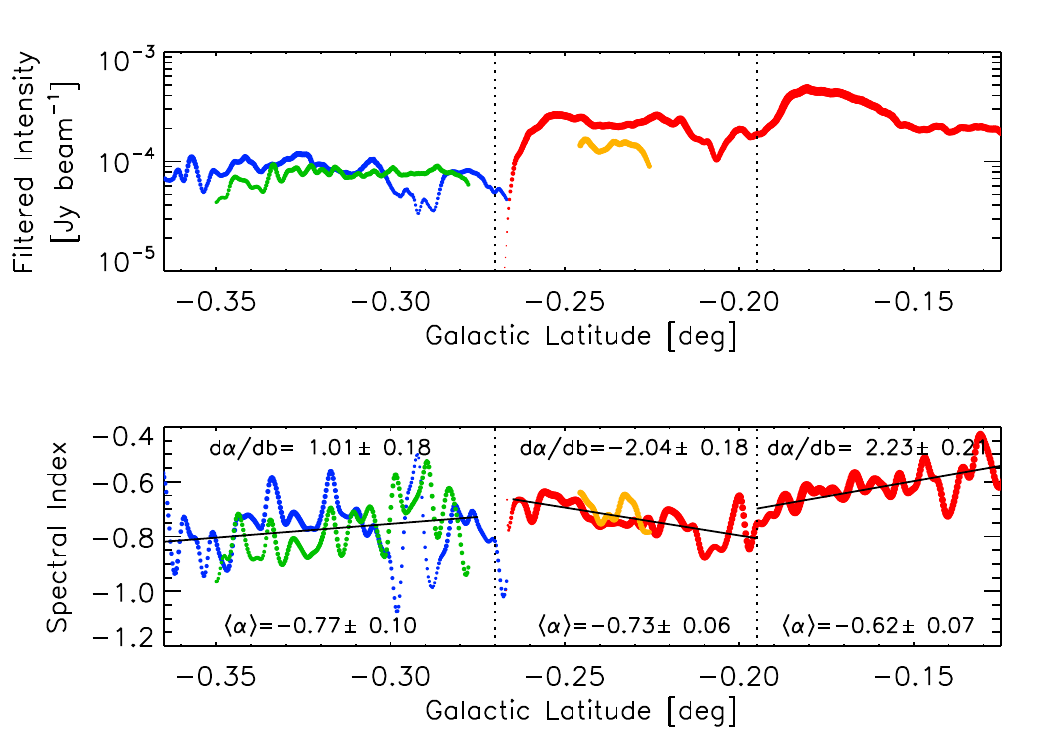}
\caption{
{\it (Top a)}
The values of the total intensity as a function of latitude
along a slice cut along the four  segments shown in the right panel of Figure 1  with a 
4$''\times4''$ beam.  
{\it (Bottom, b)}
Similar to the left panel, the mean spectral index values and 
gradients with their uncertainties as a function of latitude at at 1.28 GHz with an  8$'\times8'''$ beam. 
The uncertainties are derived from fitting the slopes. 
Color coding corresponds to different 
pieces of 
sub-filaments, as shown in right panels of Figure 1. 
}
\end{figure}

Total intensity images are constructed using CASA \citep{casa22} task {\it tClean} 
with gridding options of mosaic, utilizing the  multi-frequency synthesis (MFS) algorithm for 
the first Taylor expansion 
term (tt0) with $R=0.6$ for robustness  visibility weighting \citep{brig95}.
The sidelobes were cleaned with the Multi-term MFS  algorithm \citep{ccw1990,raucornwell2011}, 
{\it mtmfs}, a deconvolution option in tClean of 
CASA.
The rms noise is 4 $\mu$Jy beam$^{-1}$ and FWHM = 4.0$''\times3.5''$
(PA$=-14^{\circ}$, the angle of the  major axis of the beam, measured east of north in celestial coordinates). To determine the flux densities of 
the compact radio sources in the vicinity of the Snake  at 10 GHz,  
with less contamination from the 
extended emission, we imaged two pointing
fields taken in the C-array configuration survey at X-band, X2F16 and X2F19, with $R=-0.6$ and a filter rejecting the shorter baseline
visibility data. The two images  (X2F16.I.R-0.6.tt0 and X2F19.I.R-0.6.tt0)  were convolved to a beam of 3.2$''\times1.6''$
(PA$=-14^{\circ}$). 



\begin{table*}
\tablenum{1}
\footnotesize
\centering
\setlength{\tabcolsep}{1.7mm}
\caption{Log of VLA X-band uv datasets}
\begin{tabular}{lllcccccc}
\hline\hline \\
&\multicolumn{8}{c}{\underline{~~~~~~~~~~~~~~~~~~~~~~~~~~~~~~~~~~~~~~~~~~~~~~~~~~~~~~~~~~~~~~~~~~~~~~~~~~~~}}\\
&{Field ID}&
{Array}&
{Band}&
{$\nu$}&{$\Delta\nu$~~~} &
{$\Delta t$}&
{Obs. time} &
{~~~~~~~~Epoch~~~~~~~~} \\
&{} &
{} &
{} &
{(GHz)} &   
{(GHz)} &
{(sec)} &
{(min)}&
{(date)}
\\
&{(1)}&{(2)}&{(3)}&{(4)}&{(5)}&{(6)}&{(7)}&(8)\\
\hline \\   
\vspace{5pt}
&\multicolumn{7}{l}{\underline{Project-ID: 21A-000 ~~~ }}& \\
\vspace{1pt}
&F04: G359.12-0.27 &C &X$^\dagger$& 10&4&3&23&July 31, 2021\\
\vspace{1pt}&F15: G359.12-0.27 &C &X& 10&4&3&23&August 16, 2021\\
\vspace{1pt} 
&F16: G359.13-0.20 &C &X& 10&4&3&23&August 16, 2021\\
\vspace{1pt}
&F17: G359.21-0.10 
&C &X& 10&4&3&23&August 16, 2021\\
\vspace{1pt} 
&F18: G359.22-0.14 &C &X& 10&4&3&23&August 16, 2021\\ 
\vspace{1pt}
&F19: G359.22-0.19 &C &X& 10&4&3&23&August 16, 2021\\
\vspace{1pt}\\
\vspace{5pt}
\\
\hline
\end{tabular}
\vspace{2pt}
\begin{tabular}{p{0.85\textwidth}}
{
For UV data --
(1) The poining ID of the survey fields.
(2) The VLA array configuration code. C array corresponds to the maximum baseline of 3 km.
(3) The VLA band code; "X" stand for the VLA bands in the ranges of
$8-12$ GHz
(https://science.nrao.edu/facilities/vla/docs/manuals/oss2013B/performance/bands).
(4) The observing frequencies at the observing band center.
(5) The total bandwidth.
(6) The integration time.
(7) The total on-source observing time.
(8) The observing date.}\\
{$^\dagger$Correlator setup: 64 channels in each of 32 subbands with channel width of 2 MHz for RR, RL, LR, LL.}\\
{
}
\end{tabular}
\end{table*}

\subsubsection{L-band data} 

The Snake was also observed with the VLA on 
March 6, and  8, 2021 in the A-array configuration 
with a correlator setup configuring 128 channels in each of 12 subbands with channel width of 1 MHz for RR, RL, LR and LL correlations.
A data calibration procedure similar to that for the X-band data was
followed concerning pipeline reduction and residual-error corrections.

Other  archival VLA observations were also  obtained  toward  Sgr C  centered at 
(J2000 17$^{\rm h}$44$^{\rm m}$35$\fs$0, $-29^\circ 
29'\,00\farcs0$) 
having  a field of view that included  the Snake. 
These A-array measurements  were  carried out at L-band (1–2 GHz) on August 20,  2015.  
The initial flagging and reference calibration was performed using the VLA casa pipeline8 and
processed with the wsclean multi-scale clean algorithm with a resolution of $\sim 1''\times2.5''$. 
Further details on the use of the VLA data can be found in \cite{heywood22}.

\subsubsection{C-band data}

The Snake was also observed at 6 GHz in the VLA  A-array configuration on July 12, 2020 with a 
wideband correlator setup similar to the X-band survey described in Table 1, i.e., with 64 channels 
in each of 32 sub-bands with a channel width of 2 MHz for RR, RL, LR, LL correlations.
The C-band data were initially calibrated using a VLA pipeline program.  
The residual errors were corrected using the J1744-03116 data and a non-point source model developed
following the procedure described in  \citep{zhao19}.

\begin{figure}
\centering
\includegraphics[scale=0.45,angle=0]{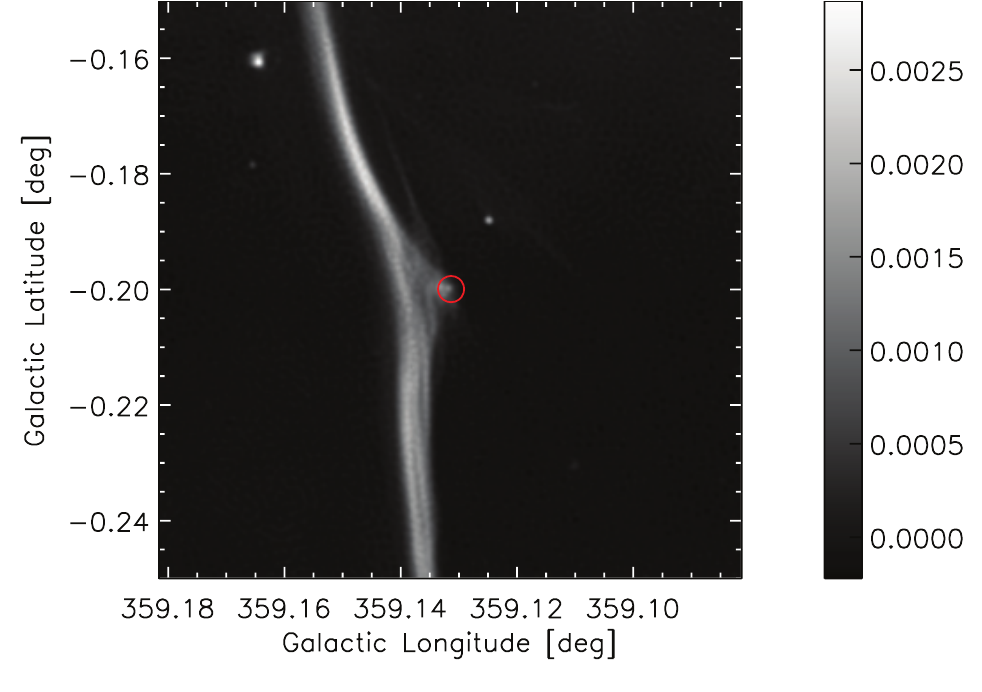}
\hspace{0.5in}
\includegraphics[scale=0.45,angle=0]{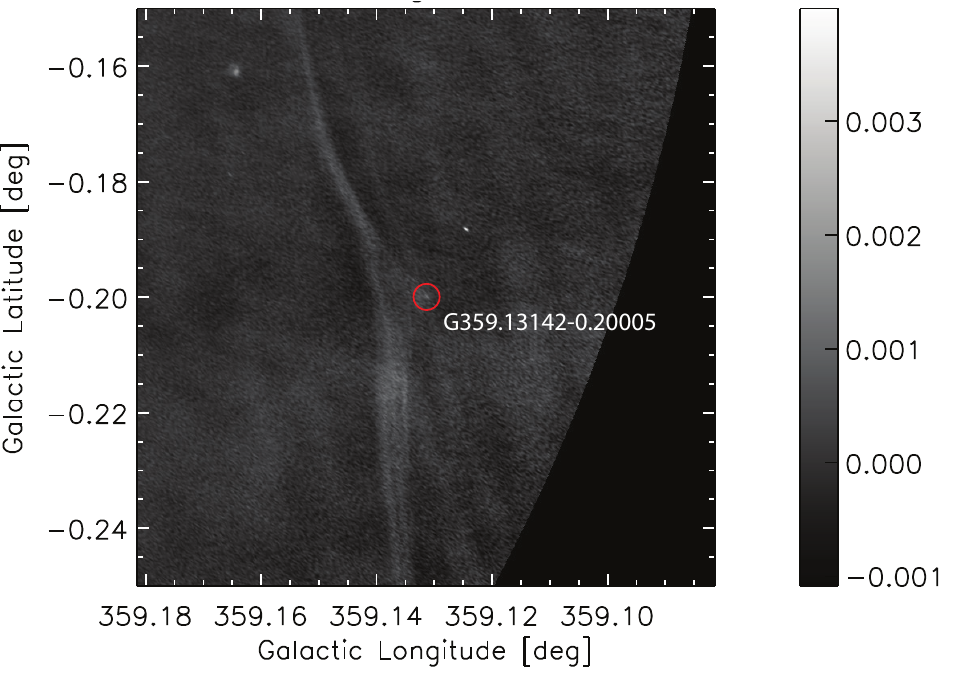}
\includegraphics[scale=0.45,angle=0]{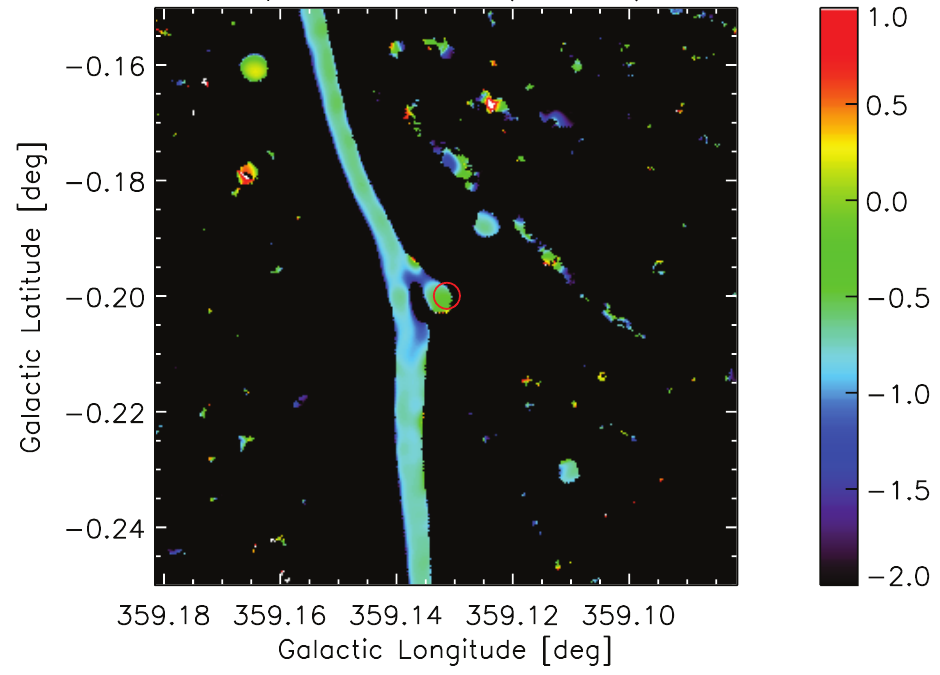}
\hspace{0.5in}
\includegraphics[scale=0.45,angle=0]{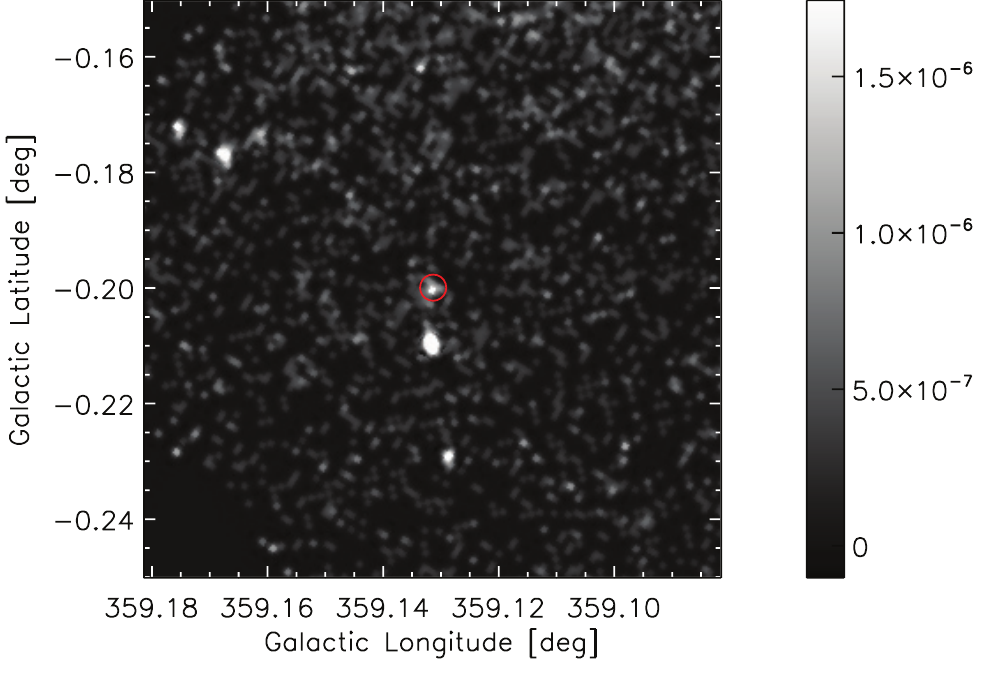}
\caption{
{\it} 
(Top Left, a), 
An unfiltered  MeerKAT image of the major kink of the Snake with a 4$''$ spatial resolution at 1.28 GHz \citep{heywood22}. 
(Top Right, b) 
Similar to (a) except a VLA image with  a spatial resolution of 6.9$''\times2.78''$ (PA=$169^\circ$) at 1.52 GHz.  
(Bottom Left, c)
Similar to the left panel  except the spectral index distribution with a spatial resolution of $\sim8''$ \citep{heywood22,zadeh22a}. 
(Bottom Right, d)
An X-ray 2-8 keV image based on Chandra observations convolved with a 2D Gaussian $4''$. 
The gray scale  bar units  in (a) and (b) are Jy beam$^{-1}$, (c)  shows the spectral index values, and in (d) counts s$^{-1}$ cm$^{-2}$.
Red circles show the location of the compact radio source G359.13142-0.20005.
}
\end{figure}

The effective total bandwidth of 4 GHz was used, centered at 6 GHz. By synthesizing a 4 GHz bandwidth 
of the wideband spectral data, we made a high resolution image with a robustness parameter of 0 for 
visibility weighting \citep{brig95} 
to have a FWHM beam of $0.69''\times0.28''$. A compact radio source
G359.13142-0.20005 was detected with a flux density  $0.20 \pm 0.01$  mJy at 6 GHz, located near 
the X-ray source. In order to determine the spectrum of the compact radio source, a continuum spectral 
image cube was constructed by averaging every four sub-band data sequentially along the frequency axis. Then, the sidelobes of the image cube were 
cleaned with the mtmfs algorithm in tClean of CASA. 
Thus, in addition to the  4-GHz bandwidth synthesized image at 6 GHz, we have eight more spectral images 
at 4.23, 4.74, 5.35, 5.76, 6.23, 6.74, 7.25, and 7.76 GHz across the C band.
A range of rms  from 0.014 to 0.025 mJy beam$^{-1}$ was achieved for the C-band spectral images.

\subsection{Radio MeerKAT  data}

Details of MeerKAT L-band observations,  mosaicking of multiple fields  and in-band spectral index measurements based on unfiltered and filtered 
images are found in 
\cite{heywood19,heywood22,zadeh22a}.
Briefly,  the L-band (856$-$1712~MHz) centered at 1.28 GHz system was used, with
the correlator configured to deliver 4,096 frequency channels, which were averaged by a factor of 4 prior to
processing and was imaged with  an angular resolution of 4$''$. 
The MeerKAT image was  spatially filtered to  remove large scale backgrounds \citep{zadeh22a}. 
Point sources in the  filtered image correspond to the spatial resolution having   FWHM $\sim 6.4''$.
An in-band spectral index mosaic was produced with  a resolution of $8''\times8''$. 
The in-band spectral index is determined using cubes of 16 filtered frequency channels.
To improve the signal-to-noise ratio (S/N), the median spectral index of each filament is determined from the median of
the pixel spectral indices along its length. 
The statistical uncertainty of the mean spectral index is typically $\sim 0.1$, but
can drop as low as $0.01$ for long bright filaments.

\subsection{X-ray  data} 

The Snake's  major kink is covered by archival Chandra ACIS-I ObsIDs 658, 2278, 2286, 18332, and 18645; and by archival Chandra HRC-I ObsID
2714. We retrieve the level=2 event files from the Chandra Data Archive. We merge these ACIS-I observations using the CIAO reproject-obs
command, and restrict the energy range for the output image to 2-8 keV. X-ray analysis of these images are discussed in \S3.2.1.

\section{Results}

We first describe the spectral index distribution along the Snake   followed by  
morphological details of the   kinks. 
Properties of  X-ray emission from a source located at the western tip of the major  kink will also be described. 

\subsection{Intensity and Spectral index distributions}

Figure 1, left shows details of the intensity distribution of the Snake with the major and minor  kinks,  
at Galactic latitudes $b\sim-0.2^\circ$ 
and $\sim-0.27^\circ$ respectively,   that show  similar morphology. 
The kinks split the Snake into three segments. 
The north and middle segments are bows to the (Galactic) east with a 
tighter radius of curvature than 
the south segment which bows to the west.
The top two  segments are brighter than the south segment by roughly a factor of two.  
There are also sub-filaments in the Snake, defined as parallel strands of filaments.  
The main filament of the Snake (top segment) breaks up into 
multiple sub-filaments at the location 
G359.14-0.19, with brightness reduced by a  factor of 1.5 at 20 cm,  
thus suggesting that the Snake is disturbed in this region. 
The equipartition 
magnetic field of the north segment is  reported to be high, reaching 0.15 mG field \citep{zadeh22a}. We notice that the Snake 
consists of bundles of filaments with small spacings. Figure 1, middle  illustrates the spectral index along the Snake. 
Figure 1, right shows 
a better color representation, distinguishing  the four  sub-filaments that are plotted in Figure 2. 
The sub-filaments  are selected because they are spatially distinct in brightness  and not because of their spectral index (see also Figure 6, left 
where  two  sub-filaments, north and south of $\sim-0.27^\circ$, correspond to the red/orange, and green/blue pairs in Figures  1, right and 2).

The mean spectral index values for each 
segment of the Snake indicates  steepening of the 
mean spectral index from -0.62 to -0.73 in the direction to the south of the major  kink. 
The mean spectral index is $\sum{\alpha(b)} / N_b$, where $\alpha(b)$  is the spectral index at each pixel (at latitude $b$), and  $N_b$ is the 
total number of pixels summed.
The gradients are change in the spectral index  per degree latitude ($degree^{-1}$). 
The regions in the direction away 
from the 
northern kink show enhanced synchrotron emissivity with a spectrum that becomes steeper to a value of $\sim-0.8$ at the southern end. 
This suggests that cosmic ray particles are more recently accelerated to the north of the major  kink, than to the south of the 
minor kink.


Figure 2 (top and bottom)  shows details of the variation of the intensity and spectral index as a function of latitude using filtered 
images of the Snake at 1.28 GHz.  A slice 
is cut along the individual segments of the intensity and the spectral index images of the Snake. 
 The spectral index gradients displayed as black lines indicate flattening of the spectral index 
away from  the major   kink,  away from $\sim b=-0.19^\circ$, (dotted vertical line). 
 A mixture of flatter extended non-thermal emission along the Snake and a steeper emission from particles injected 
at the major kink is likely to explain the unusual
spectral gradient.

The linear segment to the south of the minor   kink near $\sim b=-0.27^\circ$, as shown in Figure 2 bottom,  
shows steepening of the mean spectral index $\alpha=-0.77$. The overall mean spectral index distribution of the 
Snake indicates steepening with decreasing latitudes. 
The Snake's minor  kink also shows a spectral index gradient in which the filaments 
become steeper away from the minor   kink, opposite to that seen in the major   kink.  
This suggests that the origin of the major   kink may be different than the minor kink.

\begin{figure}
\centering
\includegraphics[scale=0.475,angle=0]{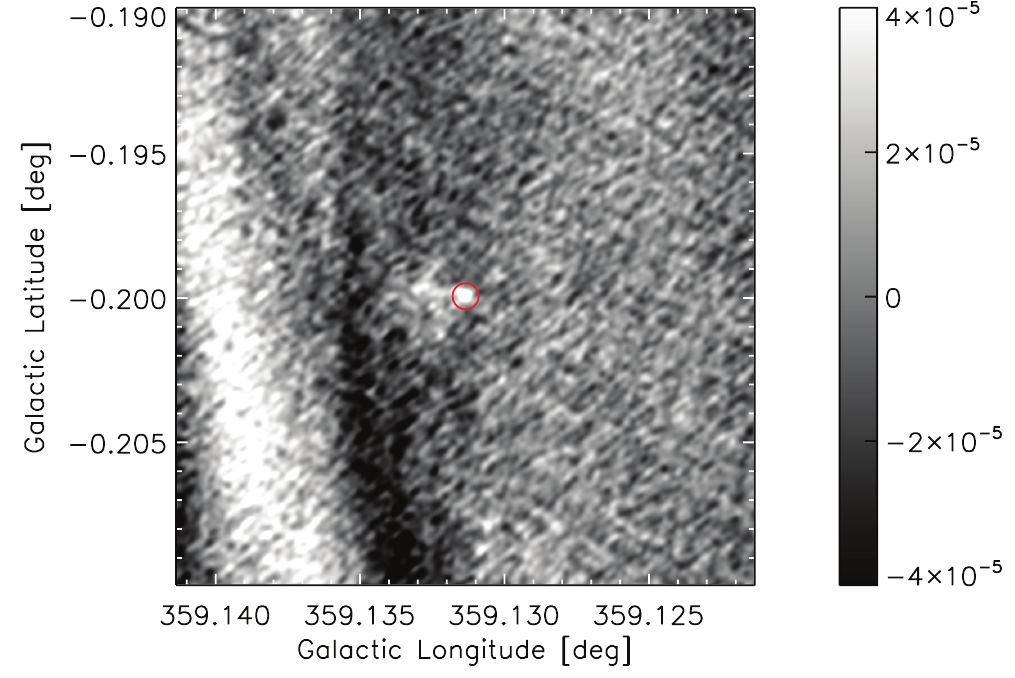}
\hspace{0.25in}
\includegraphics[scale=0.475,angle=0]{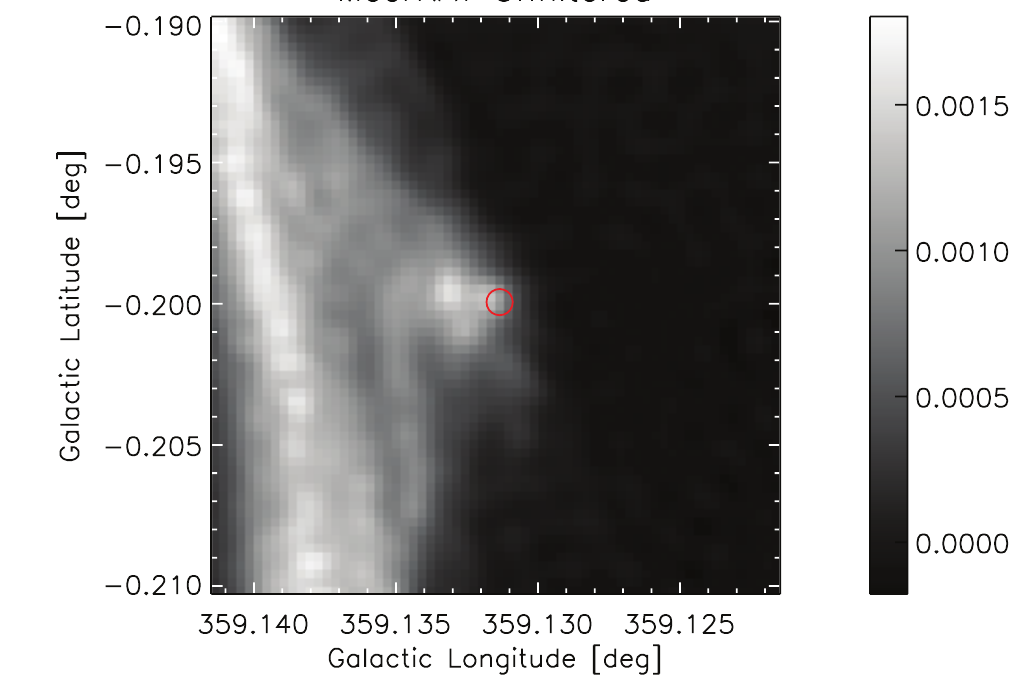}
\vspace{0.75in}
\includegraphics[scale=0.475,angle=0]{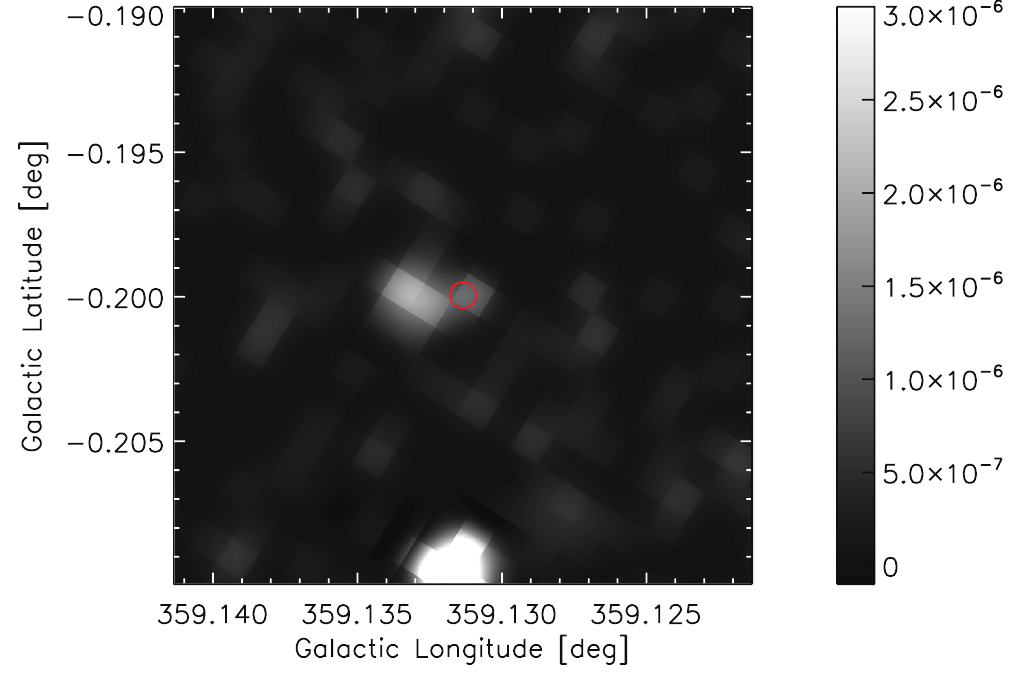}
\hspace{0.25in}
\includegraphics[scale=0.475,angle=0]{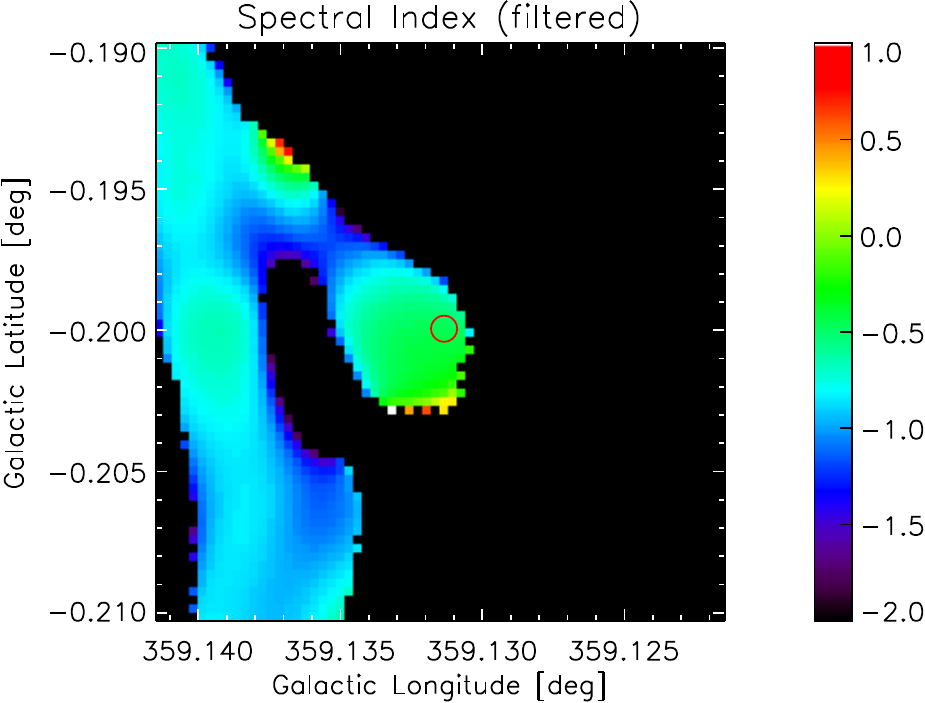}
\caption{
{\it (Top Left, a)} 
A 6 GHz VLA image of the major  kink reveals the appearance of a bow-shock morphology with a compact source 
and diffuse emission behind it with a resolution$\sim1''\times0.8''$. 
{\it (Top Right, b)} 
 Same as (a) except 
the total intensity image of the major  kink at 1.28 GHz with a resolution of $4''$. 
{\it (Bottom Left, c)} 
An X-ray counterpart to the radio sources appears to consist of two components 
coincident with the compact radio source and its extended 
liner  morphology.
{\it (Bottom Right, d)} 
The spectral index of the major kink shows a mixture of steep and flat spectral at 20cm.  
The grayscale  bars units in  (a), and (b) are in Jy beam$^{-1}$ and (c)
is in counts s$^{-1}$ cm$^{-2}$. 
} 
\end{figure}

\subsection{Major  kink G359.13-0.20}

Figures 3 (top left, top right) show close-up views of the $6'\times6'$ region of the major  kink at 1.28 and 1.52 GHz based on MeerKAT and VLA 
observations. 
A red circle shows the location where a compact radio source 
G359.13142-0.20005 
is noted to the west of the filaments.  
 The spectral index of this  source is  discussed below. 
We note 
distorted structure of the filaments, diffuse emission connecting the compact source to the major kink, 
as well
as multiple parallel sub-filaments,  with a separation of $\sim7''$, to the 
south of the major kink. 
In addition, Figure 3 (bottom left) shows a close-up view of the spectral index image of the major kink with a spatial resolution of $8''$ indicating
 that the compact source surrounded by an extended diffuse emission 
has  a flatter spectrum $\alpha\sim-0.3$ than the rest of the filaments. 
Altogether, these morphological and spectral 
characteristics 
suggest an interaction at the location of the major   kink. 

\subsubsection{X-ray analysis} 

Using a 
merged image (from the CIAO {\it reproject-obs} command) of the ACIS-I exposures in the 2-8 keV band (totaling 36.3 ks), we used {\it wavdetect} 
to find the J2000 (Galactic) coordinates of two  sources:  $17^{\rm h} 44^{\rm m} 21\fs57, -29^\circ 47' 10\farcs0\,$, 
hereafter $''$southern 
source$''$, (G359.1316-0.20160),  and  
$17^{\rm h} 44^{\rm m} 19\fs22, -29^\circ 46' 53\farcs4$, $''$ northern source$''$,  (G359.13142-0.20005).  
The southern  source is clearly 
detected in the HRC-I image and in 
ACIS images below 2 keV (indicating a foreground source), while the northern 
 source is only detected above 2 keV, consistent with a Galactic 
Bulge 
distance. 
The northern  Chandra source has no counterparts within 1$''$ in a Vizier search except for Chandra source identifications in the Chandra Source
Catalog \citep{evans10,muno09,wang16}.
The  southern  source  is consistent (within 0.2$''$, smaller than the typical Chandra ACIS uncertainty of 0.8$''$) with Gaia 
source4056860326361166848, G=19.6, geometric parallax distance estimate of 800$^{+300}_{-200}$ pc \citep{bailer-jones21}. 
 We do not consider the southern source further.

\begin{figure}
\centering
\includegraphics[scale=0.35, angle=0]{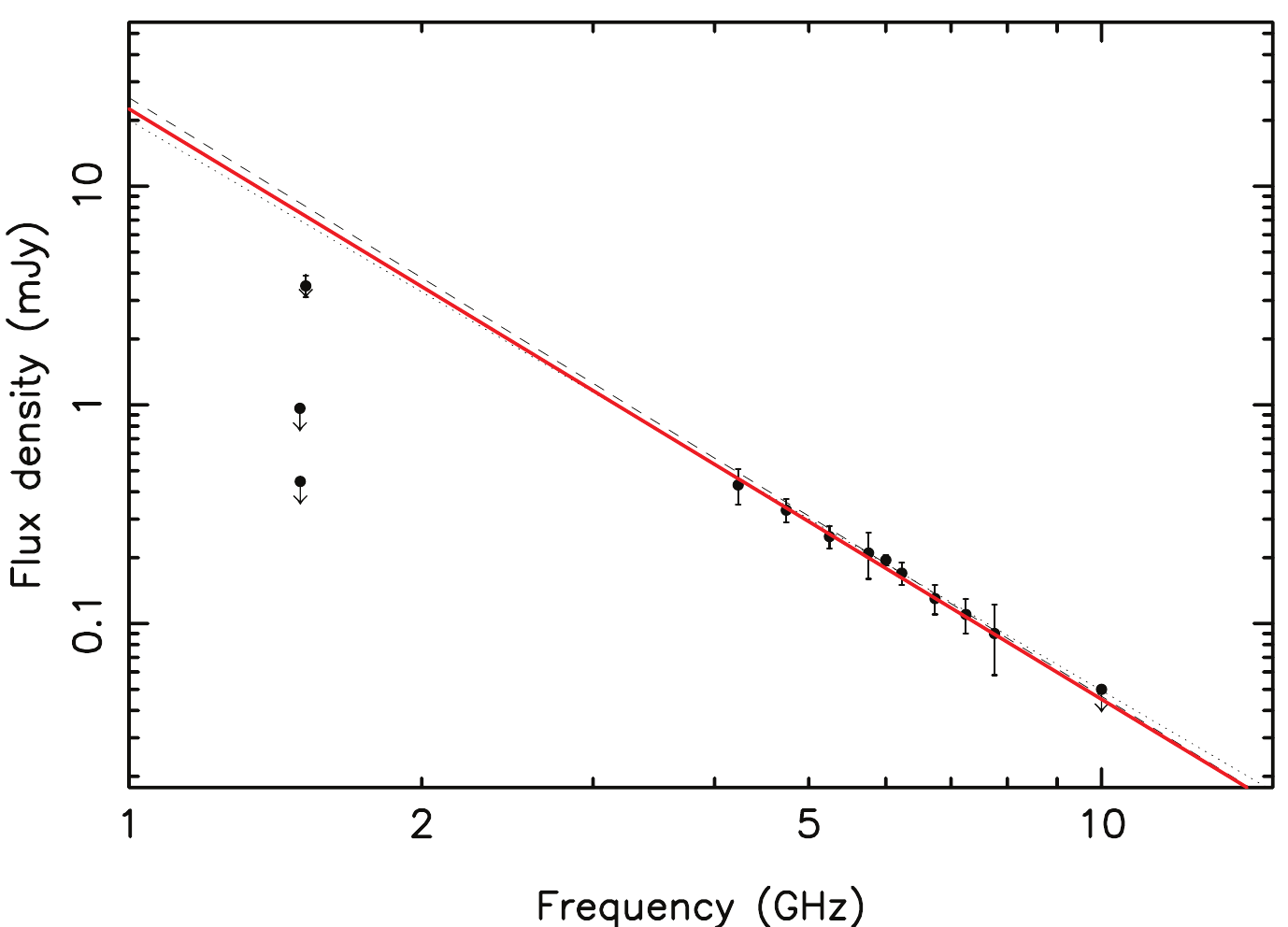}
\includegraphics[scale=0.55,angle=0]{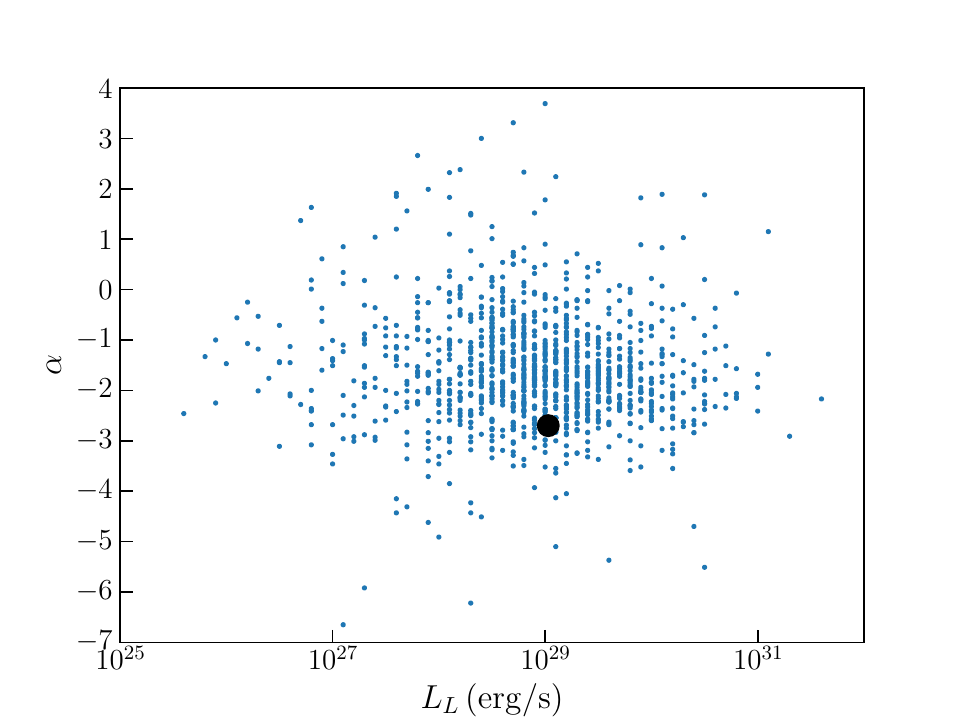}
\caption{
 {\it (Left, a)} 
The spectrum of the compact source in the major kink G359.13142-0.20005
gives  a spectral index $\alpha=-2.7$. 
The spectrum is fitted by  a power-law $F_\nu \propto \nu^\alpha$ for C-band data only (dashed line)  and all data (dotted line) listed in Table 2, 
respectively.  We derived
$\alpha$ and $\sigma_\alpha$ from a least-squares  fitting weighted by the inverse variance of the measurements, as discussed in the text. 
The red line on  the spectrum plot  shows 
the best fitted slope.
{\it (Right, b)} 
Blue points show the spectral index and L-band pseudo-luminosity for pulsars in the Thousand Pulsar Array program  \citep{posselt23}.  
The black point indicates the spectral index and  L-band luminosity of the Snake compact source.
}
\end{figure}

The  northern  X-ray source matches the position of a  compact steep spectrum  radio source
within the 0.8$''$  Chandra astrometric 
accuracy\footnote{https://cxc.harvard.edu/cal/ASPECT/celmon/}. There is a hint of extended emission from this X-ray  source, 
though it is not 
formally detected by wavdetect. In the combined X-ray image, as shown in Figure 3 (bottom right), the  point source has 13$\pm$4 counts according to {\it 
wavdetect}, while the 
extended  emission, pointing to the east (Galactic coordinates, see Fig. 4 bottom left) of the compact X-ray source has about 40 counts.

We extracted spectra from both the point source and the extended emission, using a circle of radius 2.5$''$ for the point source and a box of 
sides 
$8.8''\times12.7''$, from the three longer ObsIDs which contained more than one photon from each source (2278, 2286, and 658). 
We bin the spectra by 1 photon/bin, and fit the spectra simultaneously in XSPEC using the W-statistic \citep{cash79}  with an absorbed
power-law model (using wilm abundances, \citep{wilms20}).
 We then  fit the point source with the normalization, spectral index, and 
hydrogen column density $N_H$  free. However, the parameters are very poorly constrained, and 
tend toward rather high values of both $N_H (2\times10^{23}$ cm$^{-2}$) and spectral index (4), which we find somewhat unlikely. We thus try fits 
with one 
parameter fixed at a plausible value. If $N_H$ 
is fixed at $6\times10^{22}$ cm$^{-2}$, then the power-law photon index is 1.1$^{+1.1}_{-1.1}$. If the 
power-law photon index is fixed at 2.0, then $N_H$ is $1.2^{+1.0}_{-0.6}\times10^{23}$ cm$^{-2}$. The unabsorbed 2-10 keV flux is 
$6.6\times10^{-14}$ 
erg s$^{-1}$ cm$^{-2}$, or $8.3\times10^{-14}$, for these two models.

We followed the same process with the spectral fitting of the extended emission. Again the parameters are very poorly constrained when all three 
parameters are free. If $N_H$ 
is fixed at 6$\times10^{22}$ cm$^{-2}$, then the power-law photon index is $-0.1\pm1.1$. If the power-law photon index is fixed at 2.0, 
then $N_H=2^{+2}_{-1}\times10^{23}$ cm$^{-2}$. The unabsorbed 2-10 keV flux is roughly $1\times10^{-13}$ erg s$^{-1}$ cm$^{-2}$ for either 
scenario.

A close view of the 
radio counterpart to the X-ray source at 6 and 1.28 GHz are  displayed in Figures 4 (top left, top right), respectively.  
The high-resolution ($\sim1''\times0.8''$) 
VLA image of  the $\sim1'$ region of the major    kink (Fig. 1 top left)
shows a compact radio source with  an elongated tail-like structure to its east. 
A Gaussian fit to the compact radio source G359.13142-0.20005 at 6 GHz (Epoch date December 12, 2020)   
gives   J2000 coordinates $17^{\rm h}\, 44^{\rm m}\, 19\fs2444\pm0.0013, -29^\circ\, 46'\,  52\farcs9651\pm0.0323$ with  a spatial resolution of 
$0\farcs72\times0\farcs32$ (PA=$170^\circ.92$). 
Figure 4 (top right) displays the MeerKAT image of the same region with a resolution of 4$''\times4''$ and illustrates that the compact radio/X-ray  
sources are  embedded within the diffuse and distorted region of the major  kink. 
Figure 4 (bottom left) shows a faint X-ray source which appears to coincide with the tail of the compact  radio source shown in 
Figure 4 (top left).  We note 
that both the radio and X-ray sources are extended, with a  head-tail morphology.  Figure 4 (bottom right) shows a close-up view of the in-band spectral index of the compact and 
extended radio emission with a resolution of $\sim8''\times8''$ at 1.28 GHz using unfiltered image cube \citep{zadeh22a}.
 The spectral index varies between 0 and -0.6. 

\begin{figure}
\centering
\includegraphics[scale=0.5,angle=0]{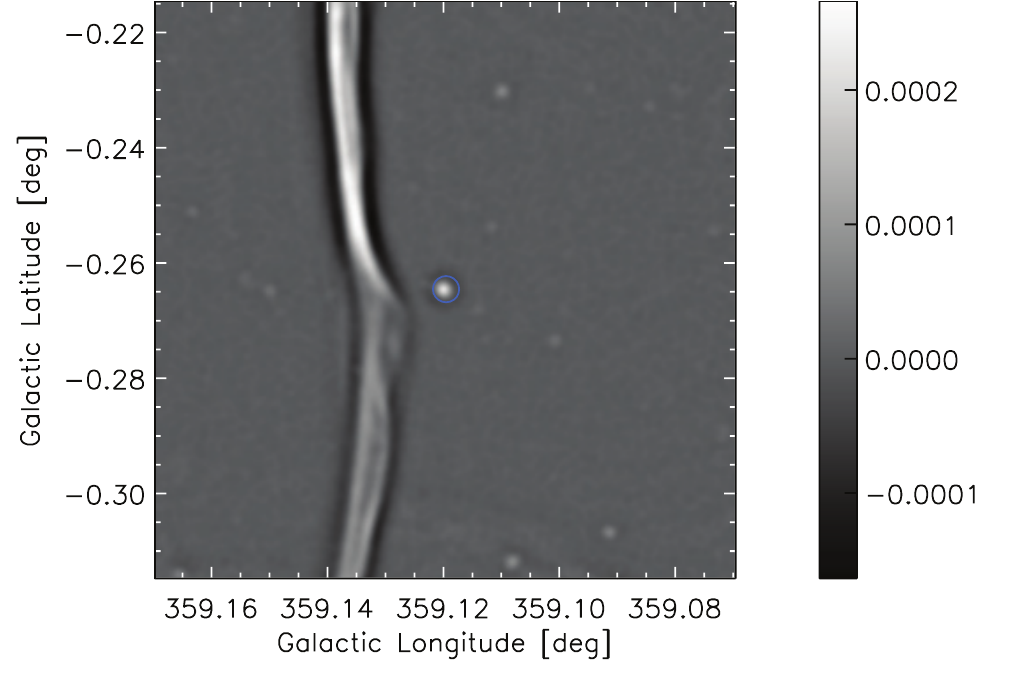}
\includegraphics[scale=0.5,angle=0]{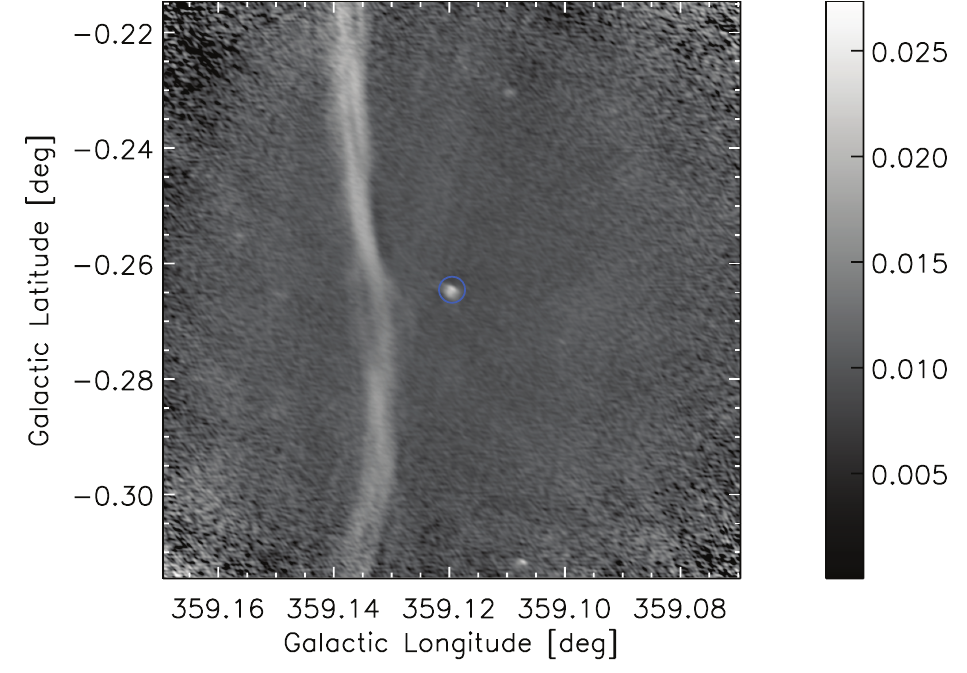}
\caption{
 {\it (Left, a)}
A close-up view of the minor kink is based on a filtered image of the  the total intensity 
 at 1.28 GHz with $6.4''$ resolution 
\citep{zadeh22a} based on MeerKAT observations. 
 {\it (Right, b)} 
Similar to Figure 3 (top right)  except close-up view of the minor kink based on a VLA image with  a resolution  6.9$''\times2.78''$ 
(PA=$169^\circ$) at 1.52 GHz.  This image is not primary beam corrected. 
The color bars units in (a) and (b) are in Jansky beam$^{-1}$. 
The blue circle corresponds to the position of the radio source G359.120-0.265.  
}
\end{figure}



\subsubsection{The compact source in the major kink G359.13142-0.20005}

To determine the nature of the compact radio source in the major kink, we used the wideband VLA C-band (6 GHz),   L-band (1.5 
GHz) and X-band (10 GHz) data sets.
With a 2D gaussian model, we fitted the compact radio source in both the wideband (4GHz-BW) synthesized image at 6 GHz and the 8 spectral
images (0.5GHz-BW)
of the image cube to determine the flux densities at the central frequencies of the sub-bands in C band from 4.23 to 7.76 GHz. The
least-squares fitting the flux-density measurements simultaneously observed in C-band on December 7, 2020 to a power-law spectrum is shown
in Figure 5, left  (the dashed straight line) with $\alpha = - 2.74 \pm 0.05$.

The X-band image was synthesized 4-GHz bandwidth data with a baseline cutoff
(> 20 kilo wavelengths ) to filter out the contamination from diffuse emission sampled in shorter baseline data providing a sensitive image
with an rms of 0.015 mJy beam$^{-1}$. The compact radio source was not detected at 10 GHz on August 16, 2021,
placing a 3-sigma upper limit of 0.05 mJy in flux density.

The L-band data at three different epochs, separated in 2 days in 2021 which spans > 5 year from  the 2015 observations. The source of 
$3.5 \pm 0.4$  mJy was detected significantly from the 2015 observation while at both epochs in the 2021 VLA A-array survey, 
the source was not
detected, giving 3-sigma upper limits of 0.9 and 0.5 mJy on March 6 and 8, 2021, respectively. The  VLA  high-resolution observations at
X-band, C-band and L-band  suggest G359.13142 - 0.20005 is variable (either due to intrinsic or interstellar scintillation) 

We note that for those flux density values measured from non-PB corrected images,
a PB correction factor A(x)  was determined from polynomial parameters given by \citep{per2016}.  The corresponding uncertainty is 
calculated using the equation (2) of \citep{zhao20}. 
      Including all the X-band, C-band and L-band data with an inverse-variance weighting, a slightly shallow spectrum of $\alpha =-2.61 \pm 0.04$ is
derived. Thus, hereafter, we take a mean value of $\alpha = - 2.7$  as the spectral index of  the compact radio source. The radio
properties of G359.13142-0.20005 are summarized in Table 2.



\begin{table*}
\footnotesize
\tablenum{2}
\setlength{\tabcolsep}{1.7mm}
\caption{The  compact radio source at 6 and 10 GHz in the major kink$^\clubsuit$}
\begin{tabular} {cccccc}
\hline\hline \\
{ID}&
{RA(J2000) ~~ Dec(J2000)}&
{$S_{\rm 6 GHz}\pm\sigma_S$}&
{$S_{\rm 10 GHz}\pm\sigma_S$}&
{$\alpha\pm\sigma_\alpha$} \\
{} &
{} &
{(mJy)} &
{(mJy)} &
{}\\
{(1)}&
{(2)}&
{(3)}&
{(4)}&
{(5)}\\
\hline \\ 
G359.13142-0.20005&17:44:19.244 $-$29:46:52.96 &0.20$\pm$0.01&$<0.05$&$-2.70\pm0.05$\\
\hline
\end{tabular}

$^\clubsuit$Column  (1) is the source ID with their Galactic coordinates, column (2) is the equatorial coordinates at Epoch J2000, with uncertainty
of $\sigma_{\rm \theta} = 0.5\theta (F_{\rm 6 GHz}/\sigma_F)$, roughly $0.02"$ or less.
Column (3) and (4)  the flux densities with 1 $\sigma$ uncertainties. For the non-detections, 3$\sigma$ upper limits are inferred. 
Column (5) the spectral index derived from least square fitting (see text).  
The errors are derived from the largest range of the 1$\sigma$ uncertainties.
\end{table*}

\subsubsection{The compact radio source, a pulsar candidate}

The steep radio spectrum ($\nu^{-2.7}$) of the compact source is suggestive of a pulsar, and the radio luminosity is consistent with this 
hypothesis.  The estimated L-band radio luminosity is $4\pi d^2 \nu S_\nu \approx 1.1\times10^{29}$\,erg\,s$^{-1}$ for $\nu = 1.2$\,GHz, $S_\nu = 
1$\,mJy  and a distance $d=8$\,kpc.  In Figure 5, right,  
we show a scatter plot of L-band spectral index  against pseudo-luminosity of the 
1200 pulsars in the Thousand Pulsar Array program 
\citep{posselt23}. The spectral index and luminosity of the 
G359.13142-0.20005 are marked by a black filled circle in Figure 5, right, and are    consistent with the general pulsar population. 


\begin{figure}
\centering
\includegraphics[scale=0.5,angle=0]{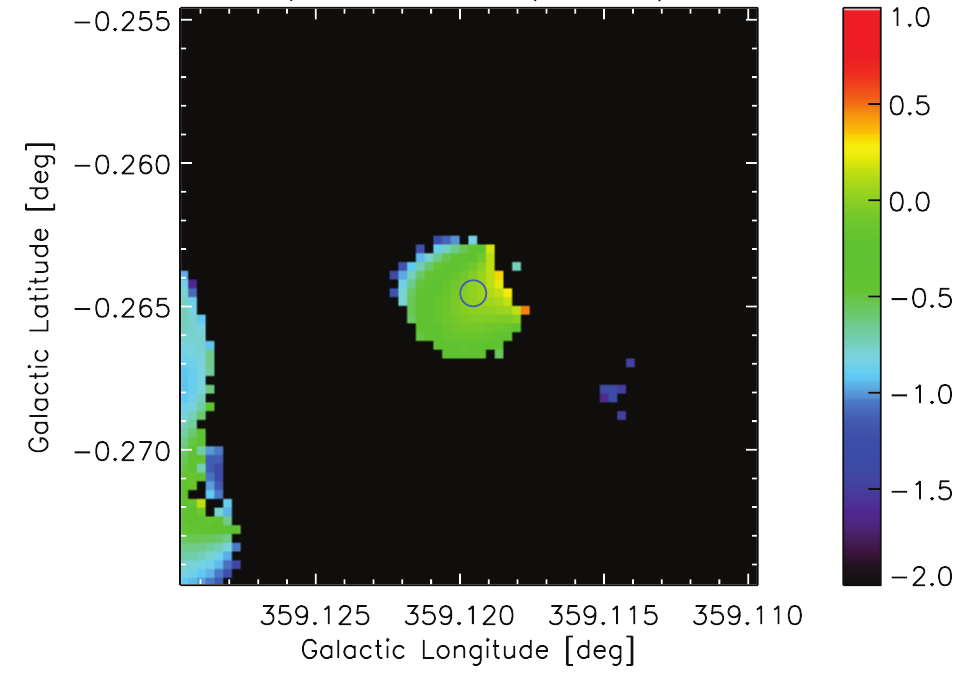}
\includegraphics[scale=0.5,angle=0]{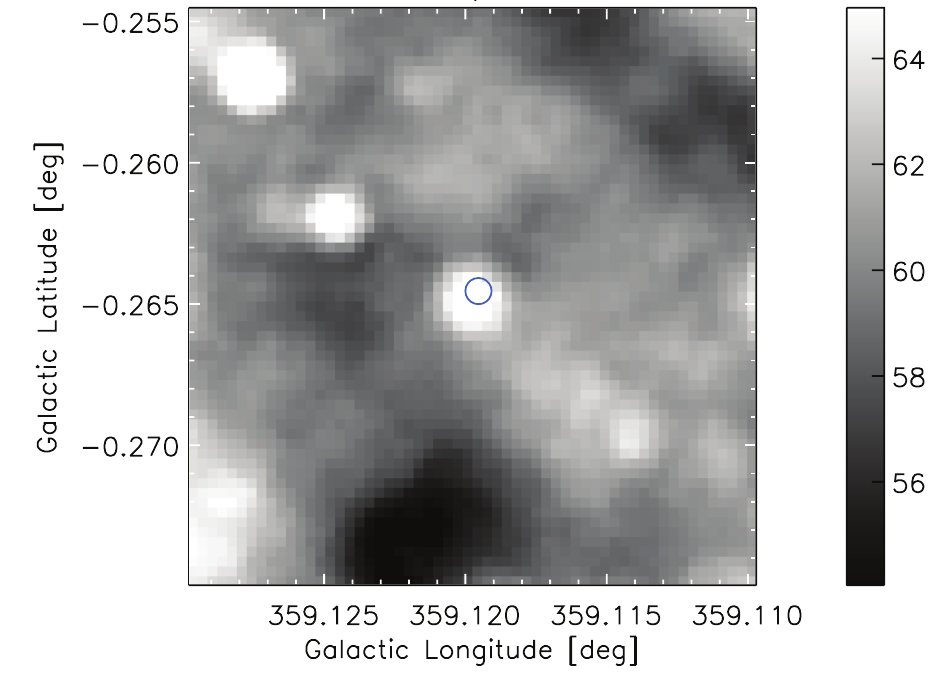}
\caption{
 {\it (Left, a)}
An in-band spectral index map of the radio source G359.120-0.265 based on filtered MeerKAT images at 20cm 
shows a spectral index near zero \citep{heywood22,zadeh22a}.  
 {\it (Right, b)} 
A 24$\mu$m infrared counterpart to G359.120-0.265 revealing thermal nature of the source. Blue circles in (a) and (b) show the location of 
G359.120-0.265. 
The color bar unit in (b) is  in MJy sr$^{-1}$. 
}
\end{figure}

\subsection{Minor kink G359.13-0.27}

Another rendition of the MeerKAT image at 1.28 GHz shows a close-up view of the minor kink in Figure 6, left.  We note bright sub-filaments to the north 
and south of the minor kink  giving the appearance of  twisted magnetized filaments.   Figure 6, right shows a higher resolution VLA image 
revealing faint diffuse emission surrounding the minor kink.   The apparent twisting of the sub-filaments is also seen  in this image. 
Morphologically, the apparent twisting  of the filaments,  sub-filamentation and the presence of diffuse emission surrounding the kink 
 are  likely  distorted by an external influence.  The variation of the mean spectral index   from $<\alpha>=-0.73$ to $-0.77$ as well as 
the mean spectral index gradient from $d\alpha/db=-1.98$ to 1.0  deg$^{-1}$ provides another support for the suggestion that there is  
disturbance at the position  of  the minor kink. 
 
Similar to the the major kink, another   candidate radio source G359.120-0.265   could be responsible for producing the minor kink. 
This radio source G359.12-0.265  
lies to the west of the  Snake, as circled in Figure 6.  Figure 7 shows the spectral index image of the compact source 
as well as 
its  24 $\mu$m infrared counterpart identified in {\it Spitzer} data. The flat spectral index close to $-0.11$ and the 
presence of 
24$\mu$m dust 
emission are consistent with  a thermal source, possible an HII region. 
This suggests that  G359.120-0.265   is unlikely to be associated 
with the minor kink.

\begin{figure}[ht]
\centering
\includegraphics[scale=1.0, angle=0]{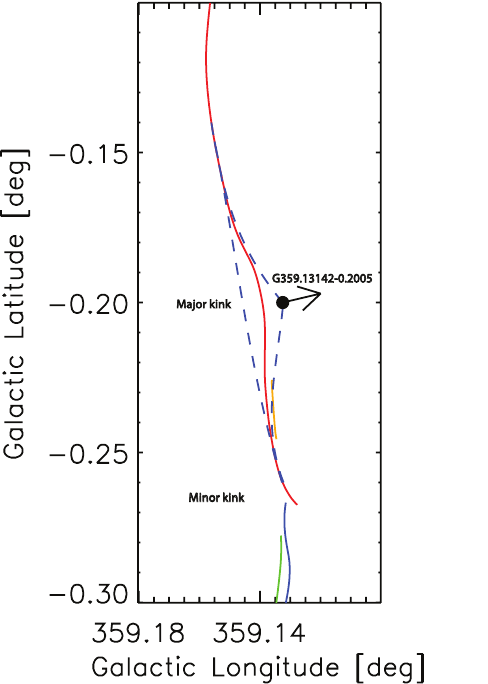}
\caption{
A schematic diagram of an object (black circle) running into the Snake and distorting the shape  of a straight filament. 
Four sub-filaments are traced  showing both kinks. Additional interpolations,  blue dashed lines,  are also drawn.  
}
\end{figure}

\section{Discussion}

The most interesting result of our study is that the Snake exhibits several signatures of an interaction at the location of the major kink: (i) the 
distortion of the Snake from a straight geometry, unlike nearly all  other filaments that show a smooth curvature; (ii) sub-filamentation 
at the kinks where the Snake is most distorted; (iii) an increase in the radio surface brightness away from the major kink; (iv) spectral index variation and 
flattening of the spectrum of the electrons away from the major kink; (v) the presence of an X-ray source coincident with the major kink as well as 
an extended tail at both radio and X-rays suggesting a bow-shock morphology. 
 

We suggest a scenario in which the major and minor kinks in the Snake result from the collision of a fast-moving nonthermal source (possibly a 
runaway pulsar) that punches through an otherwise smooth curved magnetic structure associated with a pre-existing Galactic center filament.  
Figure 8 shows a schematic diagram of this scenario where two cusps representing major and minor kinks. 
  If the interaction of the fast-moving 
object is sufficiently strong, the magnetized filament 
is pulled to one side (to the west of the vertical filament), 
creating the major kink and sending disturbances propagating along the filament in opposite directions.  If the interaction 
is sufficiently violent, the disturbance propagates as a kink-like shock.  We identify the minor kink as the current location of such a shock 
propagating southwards from the interaction point, i.e.  from the major kink.  We also note multiple sub-filaments at the location of the minor 
kink, tracing the disturbance to the south. There is, however,  no corresponding minor kink to the north, presumably because 
the natural signal speed of the magnetized filament there is significantly faster than to the south.  In this case the disturbance propagates along 
the filament as a smooth transverse wave traveling at the filament's signal speed.  This picture is supported by the observation that the 
brightest segment of the Snake lies to the north of the major kink and likely has a significantly higher magnetic field than the segment to the 
south. 

We have already suggested that the steep-spectrum and radio luminosity of the compact radio source are consistent with the time-averaged pulsed 
emission of a pulsar.  The extended radio emission visible at 6 and 1.4\,GHz has a head-tail structure, and its orientation is consistent 
with the motion required by the interaction scenario above.  This is likely synchrotron emission from electrons accelerated in a shock created 
by the interaction of a pulsar wind nebula (PWN) with the local interstellar medium, or from oppositely-directed jets of electron-positron 
pairs emerging from the poles of the pulsar and swept back by its motion through the surroundings.

We now consider the X-ray emission, which although faint and relatively poorly resolved, is consistent with a head-tail structure.  One possibility 
is that this is also synchrotron radiation from energetic 
electrons and positrons directly accelerated by the pulsar.  In support of this, we note that the compact and extended X-ray components that we detect have 
similar unabsorbed fluxes between 2 and 10\,keV, $F_x \approx 1\times10^{-13}$ erg\,cm$^{-2}$ s$^{-1}$.  At 8\,kpc this yields X-ray luminosity 
$L_x \approx 8\times10^{32}$\,erg\,s$^{-1}$, in the center of the $10^{29}$--$10^{37}$\,erg\,s$^{-1}$ range of known PWN and their associated 
pulsars (see, e.g.~ Table 1 and Fig.~1 of Hsiang \& Chang 2021).

 \section{Summary}

We presented the spectral index and morphological details of 
one of the most prominent nonthermal radio filaments, the Snake,  
in the Galactic center. 
Radio  observations   revealed a steep spectrum radio source at the location where the 
Snake is most distorted, known as the major kink. 
The steep radio spectrum ($\nu^{-2.7}$) of the compact source is suggestive of a 
fast-moving object  ($\sim500-1000$ km s$^{-1}$), likely a pulsar, that is interacting with the Snake.  
In this picture, the fast-moving object drags  material from the Snake, with a trail behind it and is 
expected to be connected to the filament by diffuse emission behind the compact source. 
To examine the interaction hypothesis, we analyzed archival Chandra data which reveal a weak X-ray source at the location of the major kink.
Alternatively, a secondary source may be responsible for disturbing
the minor kink, independent of the the pulsar candidate.
A thermal source lies to the west of the Snake, and there is no extended emission between the thermal source and
the Snake. The physical picture of the pulsar wind nebula candidate and the vertical nonthermal filaments of the Snake resembles
other filamentary structures in the Galactic enter. Two examples are
the  nonthermal compact radio and X-ray  source G0.13-0.11 lying along a nonthermal filament of the  Radio Arc \citep{churazov23} and
the compact radio source in the network of Harp filaments \citep{thomas20,heywood22,zadeh22a}. It is possible that a PWN injecting
synchrotron emitting electrons into spatially intermittent magnetic field flux tubes  \citep{thomas20}
could explain the origin of these filaments including the Snake.

Future sensitive and high-resolution  imaging at  radio and X-rays including proper motion measurements  will examine the 
interaction picture that we described here. These measurements 
provide insights on the origin of one of most remarkable radio filaments in the Galactic 
center

\section{Data Availability}

All the data including   MeerKAT  that we used here are available online and are not proprietary.
We have reduced and calibrated these data and these are available if  requested.



\section*{Acknowledgments}
This work is partially supported by the grant AST-2305857 from the
National Science Foundation. Work by R.G.A. was supported by NASA under award number 80GSFC21M0002.
The National Radio Astronomy Observatory is a facility of the National Science Foundation
operated under cooperative agreement by Associated Universities, Inc.


\vfil\eject

\bibliographystyle{mnras}

\begin{thebibliography}{}
\expandafter\ifx\csname natexlab\endcsname\relax\def\natexlab#1{#1}\fi







\bibitem[\protect\citeauthoryear{Alexander}{2012}]{alexander12} Alexander T., 2012, EPJWC,  5001, EPJWC..39

\bibitem[Alves et al.(2015)]{alves15} Alves, M.~I.~R., Calabretta, M., Davies, R.~D., et al.\ 2015, \mnras, 450, 2025 

\bibitem[\protect\citeauthoryear{Bailer-Jones et al.}{2021}]{bailer-jones21} Bailer-Jones C.~A.~L., Rybizki J., Fouesneau M., Demleitner M., 
Andrae R., 2021, AJ, 161, 147. doi:10.3847/1538-3881/abd806


\bibitem[\protect\citeauthoryear{Bicknell \& Li}{2001}]{bicknell01} Bicknell G.~V., Li J., 2001, ApJ, 548, L69

\bibitem[Boldyrev \& Yusef-Zadeh(2006)]{boldyrev06} Boldyrev, S. \& Yusef-Zadeh, F.\ 2006, ApJ, 637, L101


\bibitem[\protect\citeauthoryear{Briggs}{1995}]{brig95} Briggs, Daniel S. 1995, PhD Dissertation,
High Fidelity Deconvolution of Moderately Resolved Sources, New Mexico Institute of Mining and Technology, 
https://casa.nrao.edu/Documents/Briggs-PhD.pdf

\bibitem[\protect\citeauthoryear{The CASA Team et al.}{2022}]{casa22} The CASA Team et al 2022, PASP, 134, 114501

\bibitem[\protect\citeauthoryear{Cash}{1979}]{cash79} Cash W., 1979, ApJ, 228, 939. doi:10.1086/156922

\bibitem[\protect\citeauthoryear{Churazov et al.}{2023}]{churazov23} Churazov E., Khabibullin I., Barnouin T., Bucciantini N., Costa E., Di Gesu L., Di Marco A., 
et al., 2023, arXiv, arXiv:2312.04421. doi:10.48550/arXiv.2312.04421


\bibitem[\protect\citeauthoryear{Clark}{1980}]{bgc1980} Clark, B. G. 1980, A\&A, 89, 377


\bibitem[\protect\citeauthoryear{Conway, Cornwell \& Wilkinson}{1990}]{ccw1990} Conway, J. E., Cornwell, T. J. 
\& Wilkinson, P. N. 1990, MNRAS, 246, 490

\bibitem[Coughlin et al.(2021)]{coughlin21} Coughlin, E.~R., Nixon, C.~J., \& Ginsburg, A.\ 2021, MNRAS, 501, 1868

\bibitem[Dahlburg et al.(2002)]{dahlburg02} Dahlburg, R.~B., Einaudi, G., LaRosa, T.~N., et al.\ 2002, ApJ, 568, 220

\bibitem[\protect\citeauthoryear{Evans et al.}{2010}]{evans10} Evans I.~N., Primini F.~A., Glotfelty K.~J., Anderson C.~S., Bonaventura 
N.~R., Chen J.~C., Davis J.~E., et al., 2010, ApJS, 189, 37. doi:10.1088/0067-0049/189/1/37



\bibitem[Ferri{\`e}re(2009)]{ferriere09} Ferri{\`e}re, K.\ 2009, AA, 505, 1183.

\bibitem[\protect\citeauthoryear{Gray, et al.}{1991}]{gray91} Gray A.~D., Cram L.~E., Ekers R.~D., Goss W.~M., 1991, Nature, 353, 237
 
\bibitem[\protect\citeauthoryear{Gray, et al.}{1995}]{gray95} Gray A.~D., Nicholls J., Ekers R.~D., Cram L.~E., 1995, ApJ, 448, 164

\bibitem[\protect\citeauthoryear{Gopal-Krishna \& Biermann}{2024}]{gopal24} Gopal-Krishna, Biermann P.~L., 2024, MNRAS, 529, L135. doi:10.1093/mnrasl/slad191

\bibitem[\protect\citeauthoryear{Heywood, et al.}{2019}]{heywood19} Heywood I., et al., 2019, Natur, 573,
235

\bibitem[\protect\citeauthoryear{Heywood et al.}{2022}]{heywood22} Heywood I., Rammala I., Camilo F., Cotton W.~D.,   
Yusef-Zadeh F., Abbott T.~D., Adam R.~M., et al., 2022, ApJ, 925, 165. doi:10.3847/1538-4357/ac449a



\bibitem[\protect\citeauthoryear{H\"ogbom}{1974}]{hog1974} H\"ogbom J. A. 1974, A\&A, 15, 417

\bibitem[\protect\citeauthoryear{Hsiang \& Chang}{2021}]{hsiagn21} Hsiang J.-Y., Chang H.-K., 2021, MNRAS, 502, 390. 
doi:10.1093/mnras/stab025

\bibitem[\protect\citeauthoryear{Johnson, Dong \& Wang}{2009}]{johnson09}
Johnson S.~P., Dong H., Wang Q.~D., 2009, MNRAS, 399, 1429

\bibitem[\protect\citeauthoryear{LaRosa, et al.}{2004}]{larosa04} LaRosa T.~N., Nord M.~E., Lazio T.~J.~W., Kassim N.~E., 2004, ApJ, 607, 302

\bibitem[\protect\citeauthoryear{Law, Yusef-Zadeh \& Cotton}{2008}]{law08} Law C.~J., Yusef-Zadeh F., Cotton W.~D., 2008, ApJS, 177, 515

\bibitem[\protect\citeauthoryear{Leahy, J\"agers, \&Pooley}{1986}]{ljp1986} Leahy, J. P., J\"agers,W. J. \& Pooley, G. G. 1986, A\&A, 156, 234

\bibitem[\protect\citeauthoryear{Liszt}{1985}]{liszt85} Liszt H.~S., 1985, ApJ, 293, L65

\bibitem[\protect\citeauthoryear{Lu, Wang \& Lang}{2003}]{lu03} Lu F.~J., Wang Q.~D., Lang C.~C., 2003, AJ, 126, 319

\bibitem[\protect\citeauthoryear{Muno et al.}{2009}]{muno09} Muno M.~P., Bauer F.~E., Baganoff F.~K., Bandyopadhyay R.~M., Bower 
G.~C., Brandt W.~N., Broos P.~S., et al., 2009, ApJS, 181, 110. doi:10.1088/0067-0049/181/1/110

\bibitem[Nicholls \& Le Strange(1995)]{nicholls95} Nicholls, J. \& Le Strange, E.~T.\ 1995, ApJ, 443, 638.

\bibitem[Par{\'e} et al.(2019)]{pare19} Par{\'e}, D.~M., Lang, C.~C., Morris, M.~R., et al.\ 2019, ApJ, 884, 170


\bibitem[\protect\citeauthoryear{Perley}{2016}]{per2016} Perley, R. 2016, NRAO EVLA Memo 195

\bibitem[\protect\citeauthoryear{Ponti et al.}{2015}]{ponti15} Ponti G., Morris M.~R., Terrier R., Haberl F., Sturm R., 
Clavel M., Soldi S., et al., 2015, MNRAS, 453, 172. doi:10.1093/mnras/stv1331


\bibitem[\protect\citeauthoryear{Posselt et al.}{2023}]{posselt23} Posselt B., Karastergiou A., Johnston S., Parthasarathy A., Oswald 
L.~S., Main R.~A., Basu A., et al., 2023, MNRAS, 520, 4582. doi:10.1093/mnras/stac3383

\bibitem[\protect\citeauthoryear{Rau \& Cornwell}{2011}]{raucornwell2011} 
Rau, U. \& Conwell, T. J. 2011,AA, 532, A71



\bibitem[\protect\citeauthoryear{Rudnick et al.}{2022}]{rudnick22} Rudnick L., Bruggen M., Brunetti G., Cotton W.,
Forman W., Jones T.~W.,
Nolting C., et al., 2022, arXiv, arXiv:2206.14319

\bibitem[\protect\citeauthoryear{Rosner \& Bodo}{1996}]{rosner96} Rosner R., Bodo G., 1996, ApJ, 470, L49


\bibitem[\protect\citeauthoryear{Sakano, et al.}{2003}]{sakano03} Sakano M., Warwick R.~S., Decourchelle A., Predehl P., 2003, MNRAS, 340, 747

\bibitem[\protect\citeauthoryear{Shore \& LaRosa}{1999}]{shore99} Shore S.~N., LaRosa T.~N., 1999, ApJ, 521, 587

\bibitem[Sofue(2020)]{sofue20} Sofue, Y.\ 2020, PASJ, 72, L4

\bibitem[\protect\citeauthoryear{Thomas, Pfrommer \& En{\ss}lin}{2020}]{thomas20} Thomas T., Pfrommer C., En{\ss}lin T., 2020, ApJL, 890, L18

\bibitem[\protect\citeauthoryear{Uchida et al.}{1992}]{uchida92}Uchida, K., Morris, M., Bally, J., Pound, M. and Yusef-Zadeh, F., 1992, ApJ, 398, 128.

\bibitem[\protect\citeauthoryear{Wang et al.}{2016}]{wang16} Wang S., Liu J., Qiu Y., Bai Y., Yang H., Guo J., Zhang P., 2016, ApJS, 
224, 40. doi:10.3847/0067-0049/224/2/40



\bibitem[\protect\citeauthoryear{Wilms, Allen, \& McCray}{2000}]{wilms20} Wilms J., Allen A., McCray R., 2000, ApJ, 542, 914. 
doi:10.1086/317016

\bibitem[\protect\citeauthoryear{Yusef-Zadeh, Morris \& Chance}{1984}]{zadeh84} Yusef-Zadeh F., Morris M., Chance D., 1984, Nature, 310, 557

\bibitem[\protect\citeauthoryear{Yusef-Zadeh, et al.}{2005}]{zadeh05} Yusef-Zadeh F., Wardle M., Muno M., Law C., Pound M., 2005, AdSpR, 35, 1074

\bibitem[\protect\citeauthoryear{Yusef-Zadeh \& Wardle}{2019}]{zadeh19} Yusef-Zadeh F., Wardle M., 2019, MNRAS, 490, L1

\bibitem[Yusef-Zadeh et al.(1997)]{zadeh97} Yusef-Zadeh, F., Wardle, M., \& Parastaran, P.\ 1997, ApJ, 475, L119

\bibitem[\protect\citeauthoryear{Yusef-Zadeh et al.}{2021}]{zadeh21} Yusef-Zadeh F., Wardle M., Heinke C., Heywood I., Arendt R., Royster 
M., Cotton W., et al., 2021, MNRAS, 500, 3142. doi:10.1093/mnras/staa3257


\bibitem[Yusef-Zadeh(2003)]{zadeh03} Yusef-Zadeh, F.\ 2003, ApJ, 598, 325. doi:10.1086/378715


\bibitem[\protect\citeauthoryear{Yusef-Zadeh et al.}{2022a}]{zadeh22a} Yusef-Zadeh F., Arendt R.~G., Wardle M., Heywood I., Cotton W., Camilo F., 
2022,ApJL, 925,  L18. doi:10.3847/2041-8213/ac4802

\bibitem[\protect\citeauthoryear{Yusef-Zadeh et al.}{2022b}]{zadeh22b} Yusef-Zadeh F., Arendt R.~G., Wardle M., Heywood I., Cotton W., 
2022, MNRAS, 517, 294. doi:10.1093/mnras/stac2415

\bibitem[\protect\citeauthoryear{Zhang, et al.}{2014}]{zhang14} Zhang S., et al., 2014, ApJ, 784, 6


\bibitem[\protect\citeauthoryear{Zhao, Morris, \& Goss}{2019}]{zhao19} Zhao J.-H., Morris M.~R., Goss W.~M., 2019, ApJ, 875, 134. 
doi:10.3847/1538-4357/ab11c4

\bibitem[\protect\citeauthoryear{Zhao et al.} {2024}]{zmyr2023} 
Zhao, Jun-Hui, et al. 2024, in preparation

\bibitem[\protect\citeauthoryear{Zhao, Morris, \& Goss}{2020}]{zhao20} Zhao J.-H., Morris M.~R., Goss W.~M., 2020, ApJ, 905, 173. 
doi:10.3847/1538-4357/abc75e

\end{thebibliography}

\end{document}